%% file: dfgpc_icse2024.tex
\documentclass[sigconf,numbers]{acmart}

\copyrightyear{2024}
\acmYear{2024}
\setcopyright{acmlicensed}\acmConference[IDE '24]{2024 First IDE Workshop}{April 20, 2024}{Lisbon, Portugal}
\acmBooktitle{2024 First IDE Workshop (IDE '24), April 20, 2024, Lisbon, Portugal}
\acmDOI{10.1145/3643796.3648450}
\acmISBN{979-8-4007-0580-9/24/04}

\acmConference[ICSE 2024]{46th International Conference on Software Engineering}{April 2024}{Lisbon, Portugal}

\usepackage{algorithmic}
\usepackage{graphicx}
\usepackage{textcomp}
\usepackage{xcolor}
\usepackage{xspace}
\usepackage{hyperref}
\include{moonen-tricks}

\hypersetup{
    colorlinks,
    linkcolor={red!50!black},
    citecolor={blue!50!black},
    urlcolor={blue!80!black}
}
\def\BibTeX{{\rm B\kern-.05em{\sc i\kern-.025em b}\kern-.08em
    T\kern-.1667em\lower.7ex\hbox{E}\kern-.125emX}}
\newcommand{\mimesis}[0]{\textsc{Mimesis}\xspace}
\begin{document}

\title{Recording and Interpreting Developer Behaviour in Programming Tasks\\
}


\author{1\textsuperscript{st} Martin Schröer}
\affiliation{%
\institution{University of Bremen}
\department{Department of Computer Science}
\city{Bremen}
\country{Germany}}
\email{schroeer@uni-bremen.de}
\author{2\textsuperscript{nd} Rainer Koschke}
\affiliation{%
\institution{University of Bremen}
\department{Department of Computer Science}
\city{Bremen}
\country{Germany}}
\email{koschke@uni-bremen.de}


\begin{abstract}
\input{abstract}
\end{abstract}



\maketitle


\input{introduction}

\input{related}

\input{study}

\input{conclusion}

\input{acknowledgements}

\bibliographystyle{plainnat}
\bibliography{dfgpc_icse2024}

\end{document}

%% file: moonen-tricks.tex
\newcommand{\bibfontsize}{\footnotesize}

\makeatletter

\let\store@thebibliography\thebibliography
\def\thebibliography#1{%
  \bibfontsize\store@thebibliography{#1}%
  \setlength{\itemsep}{0pt}%
  \setlength{\parsep}{0pt}%
}
\let\store@caption\caption
\def\caption#1{\leavevmode\vspace*{-1.2ex}\store@caption{#1}\vspace*{-.8ex}}

\renewcommand\paragraph{\@startsection{paragraph}{4}{\z@}%
{0.7ex \@plus1ex \@minus.2ex}{-1em}{\normalfont\normalsize\bfseries}}
\makeatother

%% file: abstract.tex
To evaluate how developers perform differently in solving programming
tasks, i.e., which actions and behaviours are more beneficial to them
than others and if there are any specific strategies and behaviours
that may indicate good versus poor understanding of the task and
program given to them, we used the MIMESIS
plug-in~\citep{schroeer2021} to record developers' interactions
with the IDE.
In a series of three studies we investigated the specific behaviour of developers
solving a specific programming task.
We focused on which source code files they visited, how they related pieces of
code and knowledge to others and when and how successfully they performed code edits.
To cope with the variety of behaviours due to interpersonal differences such as
different level of knowledge, development style or problem solving stratiegies,
we used an abstraction of the observed behaviour, which enables for a better
comparison between different individual attributes such as skill, speed and
used stratiegies and also facilitates later automatic evaluation of behaviours,
i.e. by using a software to react to.
%
%
%
%


%% file: introduction.tex
\section{Introduction}

There have been several approaches to describe how developers perform program 
comprehension, i.e., how they tackle the issue to understand an existing program 
which they have to perform a task with, e.g., fixing a bug or introducing a new 
functionality.
Early works started with models describing how knowledge about a program is 
supposedly acquired and organised within the developers' 
mind~\citep{shneiderman1979syntactic,brooks1983towards, 
pennington1987comprehension, soloway1988knowledge, vonMayrhauser1995industrial, 
vonMayrhauser1995program}. 
Later research attempts were devoted to describe the particular strategy
developers may apply in programming tasks in general~\citep{sillito2008asking} 
and more 
specific~\citep{ko2006exploratory, grigoreanu2012strategies}.
Recent works focus more specifically on individual aspects of program 
comprehension, such as the particular knowledge developers 
need~\citep{ko2007information, maalej2014comprehension, sadowski2015developers} 
and the pecularities of different working tasks~\citep{ko2006exploratory, 
roehm2012professional} as well as the influence of experience on the different
comprehension strategies~\citep{ko2003individual}.
With the introduction of the \textit{Information Foraging 
Theory}~\citep{pirolli1999information} to the understanding of how a search for 
knowledge for program comprehension is conducted in 
particular~\citep{lawrance2008using}, new approaches to observe and understand 
the specific working steps involved have 
emerged~\citep{lawrance2013programmers}. 
Recent works building on these approaches revealed that the models and 
assumptions seem to be incomplete, as working patterns could not 
be observed~\citep{piorkowski2013whats} as predicted by corresponding 
literature.
These findings motivated us to conduct our studies.
In this article, we will discuss findings from interaction data
recorded from developers performing programming tasks and
furthermore we will also outline the usage of specific action patterns
of recorded developer behaviour to evaluate the developers performance
in comprehension and task solving.


%% file: related.tex
\section{Related Work}

\citet{roehm2012professional} observed the development activities of 
professional developers to investigate how developers comprehend software.
They found that most of the developers they observed employed a recurring 
comprehension strategy based on the context of a task, whereby the particular 
strategy chosen by an individual developer differed among them, e.g., some 
developers started with reading the code, while others preferred starting with 
inspecting the documentation or requirements.
In contrast to our study, they observed their participants using a
think-aloud method~\citep{ericsson1993protocol} and conducted interviews.

\citet{grigoreanu2012strategies} investigated the strategies and specific 
sensemaking steps which participants showed in their study while performing a 
debugging task. 
They gathered data from  think-aloud protocols~\citep{ericsson1993protocol} 
recorded from their participants and used a visualisation representing the 
participants' sensemaking steps.
Their findings were that in debugging there are multiple sensemaking loops on 
different levels of the comprehension process, each utilizing specific 
sensemaking sequences. 
This was later challenged by the findings of \citet{piorkowski2013whats}, who 
neither could observe presence of those sensemaking sequences or their assumed 
distribution in the comprehension process, nor found support for the assumption 
of a hierachical sensemaking loop.
In their work, they revealed non-linear action patterns such as "Oscillate",
i.e., developers moving forth and back between two adjacent comprehension
states and "Restart", i.e., developers advancing to the highest state
of comprehension, then returning to the lowest state and starting over.
In contrast to our work, they focused on actions of end-users debugging
\textsc{Microsoft Excel} spreadsheets based on think-aloud protocols and
visualised assumed sensemaking steps, while we recorded and visualised
interactions of developers with the IDE while performing programming tasks on a
program's source code.

\citet{lawrance2013programmers} investigated how professional developers are 
fixing bugs, with focus on the aspect how information needed to solve the task 
is sought for, inspired by ideas from the \textit{Information Foraging Theory 
(IFT)}~\citep{pirolli1999information}, suggesting that humans searching for 
information are using similar methods as a predator foraging for prey, e.g., 
following "information scents".
They found that their participants uttered far more scent-related questions 
than such relating to a hypothesis about the program they tried to understand.
Their work contrasts to ours in that they used a video recording of their
participants who were instructed to "think aloud"~\citep{ericsson1993protocol}
about what they are doing and they limited their research specifically to
debugging activities.
Also, the effectiveness of cue- or "information scent"-based strategies to 
program comprehension was investigated by \citet{piorkowski2016foraging}, who 
found in a corresponding study, that while developers often followed scents, 
their prediction about the profit and cost of such pursuits was usually rather 
bad: 50 percent of the developers' navigations were leading to less valuable 
information (if any) than they had expected, while the cost of such 
information pursuit exceeded their anticipation by nearly 40 percent on average.
In a different study, \citet{piorkowski2013whats} used video recordings to 
transcribe the development steps expressed by the recorded developers in the 
form of "think aloud 
protocols"~\citep{ericsson1993protocol}.
They then used the debugging strategy code set by 
\citet{grigoreanu2012strategies} to code the utterances of the developers they 
observed according to the taxonomy of questions that \citet{sillito2008asking} 
defined, revealing that not only the sequence, but also the distribution of 
comprehension strategies displayed did not, or did only partially match the 
corresponding predictions, such as progressing from an initial focus point to 
finally understanding groups of groups \citep{sillito2008asking}, nor was
displaying the corresponding goal patterns \citep{grigoreanu2012strategies}. 
The developers displayed "repeating" or "oscillating" goal patterns, 
rather than following a linear or hierarchical sequence that was 
predicted by literature.
In contrast to their study, we use more fine-grained interaction recordings and
a graphical visualisation of developer interactions, where their work used a
numerical representation of comprehension steps assumed from their
participants' recorded utterances.

%% file: study.tex
\section{Study}

We used the \mimesis tool~\citep{schroeer2021} to record developers'
interactions with the IDE. \mimesis captures various interactions,
among others it captures interactions such as navigations,
e.g., opening and switching between files,
code events such as selection, code changes or searching for keywords and
debugger usgage, such as setting breakpoints, starting the debugger
and hitting a set breakpoint.
\mimesis allows to run an Eclipse session from within a web browser, not
requiring any additional download on the client side.
It also offers the functionality to display instructions and a survey to
participants before they start working with the IDE.
In our studies, we used the source code of
JabRef\footnote{\url{%
https://www.jabref.org%
JabRef}},
a well known scientific literature reference manager, thus providing an
authentic code base to have participants work with.

\subsection{Study Design}
For the conduction of the studies we chose scenarios in which developers
had to perform a rather small, yet not trivial edit, thus allowing us to
collect interaction data which was then investigated for the presence of
observations reported by related works~\citep{sillito2008asking,
grigoreanu2012strategies, roehm2012professional, piorkowski2013whats,
minelli2014visualizing, minelli2015know}.
Specifically, we wanted to examine the recordings for the presence of
rather linear or hierarchical working patterns indicating
the comprehension steps as suggested by the model of
\citet{sillito2008asking}, the application of comprehension strategies
as described by \citet{grigoreanu2012strategies}, and the specific
action patterns that were observed by \citet{piorkowski2013whats}.

Three independent studies were conducted:
A) Participants were asked to add a missing "Save" button next to the
"New Library" and "Open Library" buttons that can be found below the
upper menubar's "File" option.
B) We introduced a bug, which led to clicking on the "Copy Version"
button in the "About" dialog of JabRef not working as expected.
The task was to fix this bug.
C) This study was based on the task given in B), but additionally,
participants were presented with a history of their previous navigations
through the code and suggested bookmarks for possibly interesting
code locations.
Participants of any of the studies descibed above were presented with a
survey asking for demographic information such as age and occupation and
their personal knowledge about programming before starting to work on
the task itself.

Previous studies have suggested that developers tend to neglect
certain task-relevant information given to
them~\citep{guzzi2011bookmarks, roehm2012professional, edmundson2013empirical,
bazrafshan2014effect, oliveira2014security}.
To investigate what kind of information provided is actually used by the
developers we used a particular file structure
that was providing useful information together with the program's
source code such as the task's instruction and additional information
in text files separate from the source code itself. The recordings
would allow us to see whether and when these files were accessed.
We distributed this information to seperate files to observe the
information utilisation behaviour of the developers, i.e., if and when they
access the information provided and what are the preceeding and subsequent
actions that they perform.

Furthermore, subjects were informed at the beginning of the trial that they
were allowed to arbitrarily decide when they have completed the task.
That is, if they assumed they had performed all necessary code changes, they
were free to independently finish their participation.
Beyond this, there was also no time limit given, neither were participants
instructed to use a particular approach to solve the task. They were
suggested to perform in whatever way they felt the most comfortable to do so.

This was done to allow the observation of the most natural, yet individual
development task solving behaviour possible and led to the participants
showing a high variety in how they performed the task, which we will discuss
in detail in the next section.

\section{Results}
\label{sec:results}

Out of over 300 participation attempts, N=51 recordings were selected for
further investigation, due to the participation being successfully finished
($n_s$ = 32), or unsuccsessful ($n_u$ = 19), but of appropriate length and showing
a reasonable attempt to solve the given task, i.e., browsing the source code and
searching for an entry point as opposed to giving up after reading the
instructions.

Overall, participants showed a broad variety in how they tried to solve the task.
The time spent on solving the task ranged from under 10 minutes 
to up to almost 2 hours. 
Beyond that, participants also showed different approaches to gain insight
into the source code. Only a part of them used the debugging functionality
($n_d=14$), whereas others used strategies such as "poor man's debugger", i.e.,
adding custom text output to certain places within the program, while others
seemed to use none of those strategies at all and just read the source code.

\subsection{Understanding Developer Behaviour}

When we used \mimesis visualisation of the recordings, we came to notice that in
spite of the broad variety between participants, there overall seem to be
distinct "phases" in participtions that could be described as follows:
1) "Investigation"--in this phase rather much navigation
    behaviour happens, i.e., file browsing and scrolling activity.
    Only few code interactions happen, if any.
2)  "Edit"--developers in this phase spend longer times
    navigating in certain code "blocks" and performing code edits, also
    sometimes use program launches or debugging, before advancing to
    another code block.
3) "Validation Phase"--again more broadly navigation behaviour
    occurs in this phase. Developers appear to have completed all essential
    code edits, they seem to be again scanning the code (similar to the
    investigation phase), probably to check whether they have considered every
    code line important to the task. This phase usually ends with some
    program launches to check the program's runtime behaviour.
From this overall observation, at first there seem to be some accordance to the
assumptions of linear or hierarchical comprehension models and
strategies~\citep{sillito2008asking, grigoreanu2012strategies}:
Developers initially seem to locate an appropriate starting point, then read the
surrounding code to expand their knowledge, perform code edits and validate
their edits at last.
But when furthermore examined in detail, it seems that developers probably
rather use a kind of recursive approach.
They may identify a suitable focus point to start from and then, by
investigating the relations of this focus point with the program expand their
knowledge, so far in according to corresponding assumptions.
But in contrast to the assumptions, developers do not seem to necessarily just
advance further in the model, but rather tend to restart, i.e., using the
partial knowledge gained by selecting and expanding a focus point to find a
better point to start over.
%

Recordings overall revealed distinct action patterns, that were similar to the
action sequences that \citet{piorkowski2013whats} observed.
We were able to observe developers scrolling forth and back
through the code, sometimes repeatedly, while occasionally switching between
the source code file and the provided documentation.
This may indicate an action pattern similar to what \citet{piorkowski2013whats}
found to be an "Oscillate" pattern, as scrolling forth and back may be
considered to be comprehending control or data flow (detailed understanding strategy),
while advancing through the code may be considered to be a code inspection
(coarse overall understanding strategy).
The behaviour of developers reading the source code file, performing some
edits, then switching back to the instruction file before returning to the code
file to advance reading and editing may be compared to the "Restart" pattern.


Regarding the use of provided additional information, all participants made
use of the provided files to some extent.
Interestingly, this also seems to correlate with the working phase they were
currently in, as all subjects accessed and read the files containing the
instructions, notes and temperature conversion information at the beginning of
the task and most of them revisited the instruction file before completing
working on the task, most likely to check for completeness.
But also a majority revisited the instruction and note files after spending
time in the "Inspection" phase and just before entering the "Edit" phase, this
behaviour may also indicate the switch between comprehension stages, i.e.,
after gathering enough information about the program to be confident enough to
begin editing it, they were checking again what specifically they were expected
to do.

\begin{figure*}[t]
  \centering
  \vspace*{-3mm}
  \includegraphics[width=\linewidth]{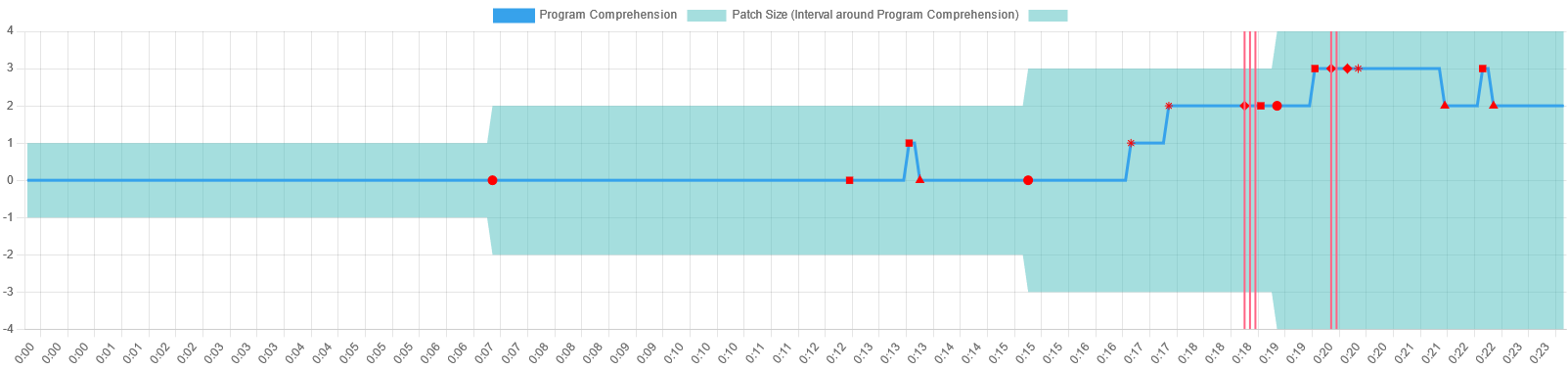}
  \vspace*{\smallskipamount}
  \includegraphics[width=\linewidth]{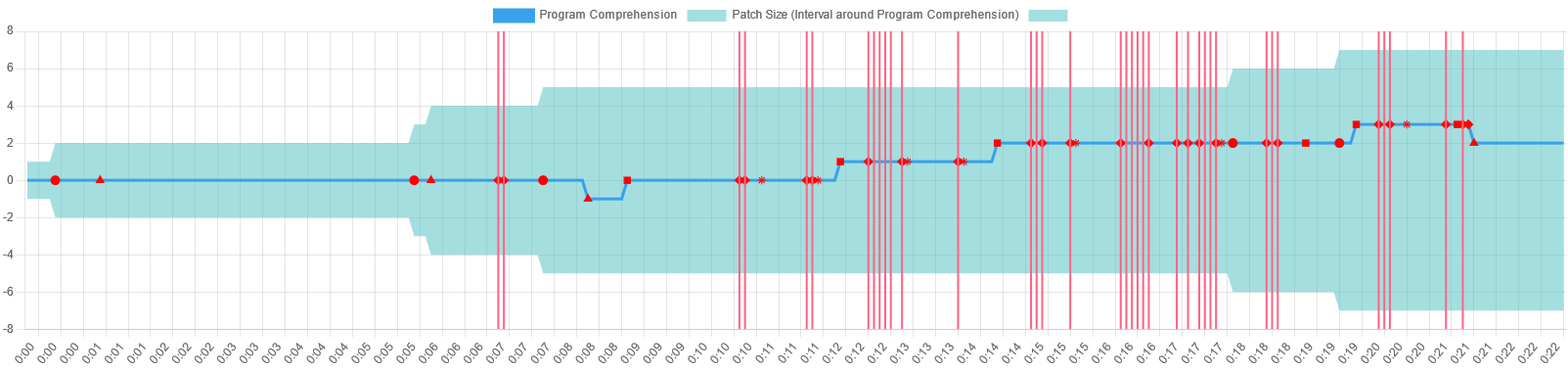}
    \vspace*{-3mm} \caption{Examples for participations}
  \vspace*{\smallskipamount}
  \textit{\footnotesize The graph shows the assumed comprehension and
  task solving performance. The surrounding area is the amount of distinct files
  visited. An ideal performance would have the graph match or even exceed the
  upper bound of this area, while overall being concise, i.e., showing only
  few action-pattern repetitions.
  Horizontal lines depict edits, symbols on the graph
  represent matched action patterns, e.g., using the debugger.\\
  In the upper visualisation, the subject at first used an alternative solution
  to the task, before realizing and revising for the actual solution.\\
  The lower visualisation displays a participation with lower overall
  comprehension, using "poor man's debugging" and a lot of additional
  unneeded changes due to uncertainty.}
  \label{fig:participations}
  \vspace*{-5pt}
\end{figure*}

\subsection{Comprehension and Task solving metric}

In an approach to evaluate developers' behaviour while also coping with the
broad variety of different problem solving strategies and specific behaviours
displayed by our participants, we developed a behaviour coding system.
Based on the developer behaviours and action patterns that could be observed
in the recordings from our aforementioned studies, we defined distinct presumed
comprehension and task solving states which are transitioned by certain developer
behaviours and action patterns.
As the navigation behaviour seems to correlate with comprehension and progress
in task solving, e.g., developers comparing new to recently vistited source files
seem to display a better comprehension
("between-patch strategies"~\citep{piorkowski2013whats}) than developers
constantly opening new files without comparing them to each other, or
constantly resorting to the task's intructions, we introduced a metric to
express this kind of navigation behaviour: Cyclissity.


$Cyclissity = 1 - P_c / N$, where $P_c$ is the position of the current file
in the file history and $N$ is the number of files visited in total.

If a developer compares two recently visited files, the resulting
cyclissity would be high, whereas visiting a file, while other files
have been visited in between would lead to a lower cyclissity value.
Visiting new files per definition results in a cyclissity of zero.
Together with corresponding action patterns, this enables for a
better understanding and evaluation of developer behaviour.

When a developer is working on a programming task, the behaviour and
specific action patterns displayed togehter with the cyclissity computed for
the corresponding file navigations lead to transitions between the aforementioned
comprehension and task solving states which are rated for increased, decreased
or not changed assumed comprehension and task solving performance.

This also allows for generalization and enables for an otherwise difficult
comparison between developers, due to different working speeds, invidiual
problem solving strategies and expertise.

When applied to the recordings of our studies, this coding scheme shows promising
results in terms of describing participants' task solving performance
(cf.~Figure~\ref{fig:participations}).

It has to be pointed out, that while a low rating can be regarded indicating
poor performance, reaching a high rating is not suffcient as a suitable
indicator for good performance alone:
While a participant repeatedly using action-patterns that are assumed to be beneficial
for comprehension and task-solving will be awarded with a high overall rating,
this may also indicate the participant performing the right steps, but still
struggling to recognise the correct code change.
For this reason, it is important to also keep track of action-pattern frequency
and repetitions, especially.

While in the successful recordings from our study, there were a number of
high scoring participants, which solved the given task using only few steps
and within short time, there also were a number of participants which,
probably due to less expertise or knowledge, showed longer time and much more
steps needed, yet still also solved the task as well.
The latter would have most likely benefitted from some kind of (tool)
assistance when--by observable repetition of certain action-patterns--showed
the indication of them struggling with the task.

%% file: conclusion.tex
\section{Conclusion}
%
%
%

In analysing IDE interactions recorded from N=51 participations of developers
performing programming tasks, we showed that in contrast to the developers
showing a wide variety in speed, knowledge and individual problem solving
strategies there seem to be general behaviours and specific action patterns
that can also be utilised to presume a developers' current quality of
comprehension and task-solving.


The specific working patterns observed in our recordings also show some 
interesting deviations from what those works suggested, especially in the 
assumption of a rather linear comprehension progress.
They rather fit to the observations of \citet{roehm2012professional, 
piorkowski2013whats, minelli2014visualizing}, in that developers seem to
recursively switch forth and back between different stages of comprehension, 
making the general assumption of a linear or hierachic progression seem to be 
too strict.
Instead,  developers seem to use some kind of refining approach, resembling to 
what \citet{sadowski2015developers} observed for general search behaviour in 
programming tasks: developers seem to start comprehending programs with a 
generic assumption or question at an almost arbitrary location within the 
program's source code, gathering more information while progressing further 
through the lines of source code, allowing them to refine their assumptions 
about the program, this leading to them restarting their inspection of 
potentially important lines, similar to what could be observed by the 
"oscillating" (scrolling back and forth) navigation behaviour in our study.

\section{Outlook}
We are currently conducting another study to validate our findings, in which
participants are asked to rate their task solving performance action-pattern
triggered times.


%% file: acknowledgements.tex
\section*{Acknowledgements}

We like to thank all study subjects for their participation.
This work was supported by the 
Deutsche Forschungsgemeinschaft (DFG, grant KO 2324/3-3).
